\documentclass[11pt]{iopart}

\usepackage[latin1]{inputenc} 

\usepackage{amssymb}
\usepackage{yfonts}

\newcommand{\f}{\frac}

\newcommand{\p}{\partial}
\newcommand{\bb}{\begin{equation}}
\newcommand{\ee}{\end{equation}}
\newcommand{\ba}{\begin{array}}
\newcommand{\ea}{\end{array}}

\newtheorem{theorem}{Theorem}

\newtheorem{corollary}{Corollary}

\newcommand{\ds}{\displaystyle }
\newcommand{\sign}{{\rm sign\,}}

\newcommand{\R}{\mathbb{R}}
\usepackage[all]{xy}
\usepackage[usenames,dvipsnames]{color}
\usepackage{graphicx}
\usepackage{caption}
\usepackage{longtable}

\usepackage{graphicx}
\usepackage{caption}

\usepackage{soul}

\begin{document}

\title[A generalised multicomponent system of Camassa-Holm-Novikov equations]{A generalised multicomponent system of Camassa-Holm-Novikov equations}

\author{}

\address{Diego Catalano Ferraioli$^1$ and Igor Leite Freire$^{2}$
 \\
 Departamento de Matem\'atica\\
Universidade Federal da Bahia, Campus de Ondina
\\ \it Av. Adhemar de Barros, S/N, Ondina, $40.170-110$ 
\\\it Salvador - Bahia - Brasil\\
 
$2$ Centro de Matem\'atica, Computa\c{c}\~ao e Cogni\c{c}\~ao\\ Universidade Federal do ABC - UFABC\\
Avenida dos Estados, $5001$, Bairro Bangu,
$09.210-580$\\\it Santo Andr\'e, SP - Brasil}
\ead{diego.catalano@ufba.br}
\ead{igor.freire@ufabc.edu.br and igor.leite.freire@gmail.com}

\begin{abstract} In this paper we introduce a two-component system, depending on a parameter $b$, which generalises the Camassa-Holm ($b=1$) and Novikov equations ($b=2$). By investigating its Lie algebra of classical and higher symmetries up to order $3$, we found that for $b\neq 2$ the system admits a $3$-dimensional algebra of point symmetries and apparently no higher symmetries, whereas for $b=2$ it has a $6$-dimensional algebra of point symmetries and also higher order symmetries. Also we provide all conservation laws, with first order characteristics, which are admitted by the system for $b=1,2$. In addition, for $b=2$, we show that the system is a particular instance of a more general system which admits an $\mathfrak{sl}(3,\mathbb{R})$-valued zero-curvature representation. Finally, we found that the system admits peakon solutions and, in particular, for $b=2$ there exist 1-peakon solutions with non-constant amplitude.

\end{abstract}

\section{Introduction}

Since the seminal paper by Camassa and Holm \cite{camassa}, hundreds works have been devoted to several aspects of nonlocal evolution equations of the Camassa-Holm (CH) type, such as: integrability, in the sense of the existence of infinite symmetries \cite{dehoho,how,mik,miknov,nov}, existence of bi-hamiltonian formulation and Lax pairs \cite{dehoho,how,nov,qiao}, and existence of (multi-) peakon solutions \cite{dehoho,how}. Further aspects of these classes of equations have also been widely investigated from many different points of view, see \cite{him1, him2, him3, holm-staley,lenjpa}. Also, some works considering systems with many components of CH type equations have been of interest, see {\it e.g.} \cite{geng,holm2010,lijnmp}.

More recently, equations with parameter-dependent nonlinearities have been considered, see for instance \cite{anco,pri1,pri2,him4,him5}, where families of equations unifying both Camassa-Holm and Novikov equations \cite{how,nov} are studied.

Motivated by the works \cite{anco,pri1,pri2}, in this paper we consider the system
\bb\label{1.1}
\begin{array}{l}
\left\{ \begin{array}{l}
m_{t}+(b+1)\,u_{x}\,v^{b-1}\,m+u^{b-1}\,v\,m_{x}=0,\\
\\
n_{t}+(b+1)\,v_{x}\,u^{b-1}\,n+v^{b-1}\,u\,n_{x}=0,
\end{array}\right.
\end{array}
\ee
where $u=u(x,t)$, $v=v(x,t)$, $m=u-u_{xx}$ and $n=v-v_{xx}$ are referred to as the momenta and $b\in\R$.

System (\ref{1.1}) is invariant under the change $(u,v)\mapsto(v,u)$. In particular, when $u=v$ the system reduces to 
\bb\label{1.2}
m_{t}+(b+1)\,u_{x}\,u^{b-1}\,m+u^{b}\,m_{x}=0.
\ee
Interestingly, equation (\ref{1.2}) reduces to CH equation for $b=1$, and to Novikov equation for $b=2$. Therefore, one may consider system (\ref{1.1}) as a two-component generalisation of both CH and Novikov equations.
Equation (\ref{1.2}) is just the equation deduced in \cite{pri1}, by using symmetry arguments and techniques introduced in \cite{ib2,ib6}. In \cite{pri2} equation (\ref{1.2}) was also re-obtained by imposing invariance under scalings and the existence of a certain multiplier (see \cite{abeu1,abeu2} for further details). Later, in \cite{anco} it was proved that (\ref{1.2}) admits peakon and multi-peakon solutions. Other properties of (\ref{1.2}) were also investigated by Himonas and Holliman in the paper \cite{him3}, by embedding it into a two-parameter family of equations.

In this paper we shall investigate system (\ref{1.1}) from several points of view. In Section \ref{sym} we compute symmetries and conservation laws. In particular, we investigate the existence of higher order, or generalised, symmetries. Then, in Section \ref{ZCR}, we show that (\ref{1.1}) can be embedded in a $4$-component system admitting an $\mathfrak{sl}(3,\R)-$valued zero-curvature representation (ZCR) which generalizes an analog result of paper \cite{Li-Liu-Pop}. In Section \ref{peak} we investigate the existence of peakon and multi-peakon solutions of (\ref{1.1}). Finally, our results are discussed in Section \ref{dis}.

\section{Symmetries and conservation laws of system $(\ref{1.1})$} \label{sym}

In this section we collect the results of the search of classical and higher symmetries of system $(\ref{1.1})$, as well as of low order conservation laws.

\subsection{Classical and higher symmetries}
By standard methods of symmetry analysis (see e.g. \cite{bk,ba,2ndbook,i,ol,vin}) one can prove the following 

\begin{theorem}\label{teo1} 
When  $b\neq 2$, the Lie algebra of classical symmetries of $(\ref{1.1})$ is $3$-dimensional with generators
\bb\label{2.1}
X_1=\f{\p}{\p x},\quad X_2=\f{\p}{\p t},\quad X_3=bt\f{\p}{\p t}-u\f{\p}{\p u}-v\f{\p}{\p v}.
\ee
On the contrary, for $b=2$, Lie algebra of classical symmetries of $(\ref{1.1})$ is $6$-dimensional, with generators $X_1$, $X_2$, $X_3$ (where $b=2$), and  

\bb\label{2.2}
X_{4}=u\f{\p}{\partial u}-v\f{\p}{\partial v},\quad X_{i}=e^{2\epsilon_{i}x}( \f{\p}{\partial x}+\epsilon_{i} u\f{\p}{\partial u}+\epsilon_{i} v\f{\p}{\partial v}),\qquad
\ee
with $i=5,6$ and $\epsilon_5=1$, $\epsilon_6=-1$.
\end{theorem}

Therefore, by computing the flows of classical symmetries admitted
by (\ref{1.1}) one gets the following 
\begin{corollary}
Under the flows $\left\{ A_{s_{i}}\right\} $ of classical symmetries
$X_{i}$, $i=1,2,...,6$, where the $s_{i}$'s denote the flow-parameters,
solutions $\left\{ u=u(x,t),\,v=v(x,t)\right\} $ of $(\ref{1.1})$ respectively
transform to:\\ \vspace{6pt}
$1)\;\left\{ u_{1}=u(x-s_{1},t),\,v_{1}=v(x-s_{1},t)\right\} $; \\ \vspace{6pt}
$2)\;\left\{ u_{2}=u(x,t-s_{2}),\,v_{1}=v(x,t-s_{2})\right\} $;\\ \vspace{6pt}
$3)\;\left\{ u_{3}=e^{s_{3}}u\left(x,te^{-bs_{3}}\right),\,v_{3}=e^{s_{3}}v\left(x,te^{-bs_{3}}\right)\right\} $;\\ \vspace{6pt}
$4)\;\left\{ u_{4}=e^{-s_{4}}u\left(x,t\right),\,v_{4}=e^{-s_{4}}v\left(x,t\right)\right\} $;\\ \vspace{6pt}
$5)\;\left\{ u_{5}=\sqrt{1+2s_{5}e^{2x}}u\left(-\frac{1}{2}\ln\left(2s_{5}+e^{-2x}\right),t\right),\,v_{5}=\sqrt{1+2s_{5}e^{2x}}v\left(-\frac{1}{2}\ln\left(2s_{5}+e^{-2x}\right),t\right)\right\} $;\\ \vspace{6pt}
$6)\;\left\{ u_{6}=\sqrt{1-2s_{6}e^{-2x}}u\left(\frac{1}{2}\ln\left(-2s_{6}+e^{2x}\right),t\right),\,v_{6}=\sqrt{1-2s_{6}e^{-2x}}v\left(\frac{1}{2}\ln\left(-2s_{6}+e^{2x}\right),t\right)\right\} $.

\end{corollary}

The special character suggested by  Theorem \ref{teo1}  for the case $b=2$ is further confirmed by the search of higher order (or generalised) symmetries. Indeed, we have not found any higher order symmetry for $b\neq 2$, whereas for $b=2$ we found the following

\begin{theorem}\label{teo2} 
When $b=2$ system $(\ref{1.1})$ admits higher order symmetries. Moreover, up to order $3$, higher symmetries in evolutionary form 
$$
Y =\sum_{|\sigma|\geq0}D_{\sigma}(\phi)\f{\p}{\partial u_{\sigma}}+\sum_{|\sigma|\geq0}D_{\sigma}(\psi)\f{\p}{\partial v_{\sigma}}
$$
are described by the characteristics (or generating functions) $Q=(\phi,\psi)$:

$$
\phi={\displaystyle \sum_{i=1}^{9}}c_{i}\phi_{i}\qquad\psi={\displaystyle \sum_{i=1}^{9}}c_{i}\psi_{i}
$$
where $Q_i=(\phi_{i},\psi_{i})$, for $i=1,...,6$, are the characteristics of classical symmetries $X_i$ (see Theorem $\ref{teo1}$), whereas for $i=7,8,9$ are given by

$$\phi_{7}=\frac{\left(u-u_{xx}\right)^{1/3}}{\left(v-v_{xx}\right)^{2/3}},\qquad\psi_{7}=\frac{\left(-v+v_{xx}\right)^{1/3}}{\left(u-u_{xx}\right)^{2/3}},
$$
\vspace{10pt}
$$
\begin{array}{l}
\phi_{8}=2\left(vv_{xx}+v_{x}^{2}-\frac{3}{2}v^{2}\right)u^{3}+2u\left(v^{2}u_{x}^{2}-u_{x}v_{t}\right)-u_{tt}\vspace{10pt}\\
\qquad+4u^{2}\left(\frac{1}{4}v^{2}u_{xx}-vv_{x}u_{x}+\frac{1}{2}v_{xt}\right),\vspace{10pt}\\
\psi_{8}=-2\left(u_{x}^{2}+uu_{xx}-\frac{3}{2}u^{2}\right)v^{3}+2v\left(-v_{x}^{2}u^{2}+u_{t}v_{x}\right)+v_{tt}\vspace{10pt}\\
\qquad+4v^{2}\left(-\frac{1}{4}u^{2}v_{xx}+uv_{x}u_{x}-\frac{1}{2}u_{xt}\right),
\end{array}
$$
\vspace{10pt}
$$
\begin{array}{l}
\phi_{9}={\displaystyle \frac{1}{2}}\left({\displaystyle v^{2}u{\it u_{t}}}-{\displaystyle v^{3}u^{2}u_{x}}\right){\it u_{xx}}+{\displaystyle \frac{1}{2}}\left(-u^{2}vv_{x}+uv^{2}u_{x}-uv_{t}\right){\it u_{tx}}\vspace{10pt}\\
\qquad+\left({\displaystyle \frac{v^{2}u_{x}^{2}}{2}}-{\displaystyle \frac{1}{2}}\left({\it v_{t}}+3uvv_{x}\right)u_{x}+u^{2}v_{x}^{2}+u^{2}vv_{xx}+uv_{tx}-{\displaystyle \frac{3}{2}}v^{2}u^{2}\right)u_{t}\vspace{10pt}\\
\qquad-{\displaystyle \frac{1}{6}}u^{3}v^{3}u_{xxx}+{\displaystyle \frac{2}{3}}u^{3}v^{3}u_{x}-{\displaystyle \frac{1}{6}}u_{ttt},\vspace{10pt}\\
\psi_{9}={\displaystyle \frac{1}{2}}\left({\displaystyle u^{2}v{\it v_{t}}}-{\displaystyle u^{3}v^{2}v_{x}}\right){\it v_{xx}}+{\displaystyle \frac{1}{2}}\left(-v^{2}uu_{x}+vu^{2}v_{x}-vu_{t}\right){\it v_{tx}}\vspace{10pt},\\
\qquad+\left({\displaystyle \frac{u^{2}v_{x}^{2}}{2}}-{\displaystyle \frac{1}{2}}\left({\it u_{t}}+3vuu_{x}\right)v_{x}+v^{2}u_{x}^{2}+v^{2}uu_{xx}+vu_{tx}-{\displaystyle \frac{3}{2}}u^{2}v^{2}\right)v_{t}\vspace{10pt}\\
\qquad-{\displaystyle \frac{1}{6}}v^{3}u^{3}v_{xxx}+{\displaystyle \frac{2}{3}}v^{3}u^{3}v_{x}-{\displaystyle \frac{1}{6}}v_{ttt}.
\end{array}
$$
In particular, the corresponding Lie algebra structure is given by the non trivial Jacobi brackets

$$
\begin{array}{l}
\{Q_1,Q_5\}=2Q_5,\quad \{Q_1,Q_6\}=-2Q_6,\quad \{Q_2,Q_3\}=2Q_2,\vspace{10pt}\\
\{Q_3,Q_8\}=-4Q_8,\quad \{Q_3,Q_9\}=-6Q_9,\quad \{Q_5,Q_6\}=-4Q_1.
\end{array}
$$
\end{theorem}

Theorem \ref{teo2} provides important indications about the property of system (\ref{1.1}) being symmetry-integrable when $b = 2$. Indeed, according to the terminology introduced in \cite{kamp} (see also \cite{fokas1987}), Theorem \ref{teo2} proves that system \ref{1.1} is {\it almost symmetry-integrable} of depth at least $3$. On the contrary, our computations up to order $3$ did not provide any higher symmetry of (\ref{1.1}) for $b\neq 2$. Thus, we conjecture that system $(\ref{1.1})$ is symmetry-integrable only in the case $b=2$.

\subsection{Conservation laws}\label{cons}

We recall that a 1-form $\Lambda=Pdx+Qdt$ is a local conservation law for (\ref{1.1}), provided that $d\Lambda\equiv0$ on the solutions of (\ref{1.1}). Local conservation laws form a real vector space and we refer the reader to \cite{abeu1,abeu2,vin} for the general theory and more details on computation techniques.   

In order to find local conservation laws of system (\ref{1.1}), one has to satisfy the following condition
$$
\begin{array}{l}
D_{t}(P)-D_{x}(Q)-\varphi\left[m_{t}+(b+1)\,u_{x}\,v^{b-1}\,m+u^{b-1}\,v\,m_{x}\right]\\
\\-\psi\left[n_{t}+(b+1)\,v_{x}\,u^{b-1}\,n+v^{b-1}\,u\,n_{x}\right]=0,
\end{array}
$$
where $(\varphi,\psi)$ are the characteristics of the corresponding conservation laws.

By performing all needed computations, for cases $b=1$ and $b=2$ with first order characteristics, we find the following

\begin{theorem}
The space of nontrivial conservation laws with first order characteristics of system $(\ref{1.1})$ is $1$-dimensional for $b=1$ and $5$-dimensional for $b=2$, with generators $\Lambda=Pdx+Qdt$ given in Table $\ref{tabl1}$.
\end{theorem}

\Table{\label{tabl1}Low order conservation laws for system (\ref{1.1}) with $b=1$ and $b=2$.}
\br
&&&\centre{2}{Components of conserved vectors}\\
\ns
&&&\crule{2}\\
$b$&$\varphi$ & $\psi$ &$P$ (density) &$Q$ (flux)\\
\mr
1 & 1 & 1 & $u+v$ &$\ba{l} -v^{2}+v_{x}^{2}-u^{2}+u_{x}^{2}+uv_{xx}\\
-uv+vu_{xx}-v_{x}u_{x}+u_{tx}+v_{tx}\ea$\\
\\

 2  & $v$  & $u$ & $uv+u_xv_x$ & $\ba{l} (vv_{xx}+v_{x}^{2}-2v^{2})u^{2}+\\
 (v_{tx}+v^{2}u_{xx}-2vu_{x}v_{x})u\\
 + vu_{tx}+v^{2}u_{x}^{2}\ea$\\
 \\
 
 2  & $-v_{x}$  & $u_{x}$ & $(u_{xx}-u)v_{x}$ & $\ba{l}-v_{x}vu_{x}^{2}+u_{x}v_{tx}\\
 +(vu_{x}v_{xx}+u_{x}v_{x}^{2}-u_{xx}vv_{x}-v_{t})u\ea$\\
 \\
 
 2  & $-v_{t}$  & $u_{t}$ & $\ba{l}((vv_{xx}+v_{x}^{2})uu_x\\+v_{tx}u_{x}+2v_{x}u^{2}v\\
 +u_{xx}v_{t}\ea$ & $\ba{l}u_{t}uv_{x}^{2}+2v_{t}vu^{2}-v_{t}vu_{x}^{2}\\+(vu_{x}u+u_{t})v_{tx}
+u_{t}uvv_{xx} \\+v_{t}(u_{x}v_{x}-u_{xx}v)u+u_{x}v_{tt}\ea$\\
\\

2  & $-(v_{x}-v)e^{2x}$  & $(u_{x}-u)e^{2x}$ & $\ba{l}\left[(v_{x}-v)(u_{x}+u_{xx})\right.\\
\left.+u(-v+v_{xx})\right]e^{2x}\ea$ & $\ba{l}\left\{ v(v-v_{x})u_{x}^{2}+\left[(v_{xx}v+v_{x}^{2}-2v^{2})u\right.\right.\\
\left.\left.+(v_{tx}-v_{t})\right]u_{x}
+v(v-v_{x})uu_{xx}\right.\\
\left.-(v_{x}^{2}+v_{xx}v-2vv_{x})u^{2}+u_{t}(v_{x}-v)\right\} \,\rme^{2x}\ea$\\
\\

2  & $-(v_{x}+v)e^{-2x}$  & $(u_{x}+u)e^{-2x}$ & $\ba{l}\left[(2v-v_x)u_{x}-uv\right.\\
\left.-(u-u_{xx})v_{x}\right]e^{-2x}\ea$ & $\ba{l}\left\{\left[uv_{x}^{2}+(v_{tx}+vv_{xx}u-2v^{2}u)\right]u_{x}\right.\\
-(v^{2}+vv_{x})u_{x}^{2}+u^{2}v_{x}^{2}\\
+uv(2u-u_{xx})v_{x}+(-vu_{tx}+uv_{tx}\\
\left.-u_{xx}v^{2}u+vv_{xx}u^{2}+v_{t}u)\right\} \rme^{-2x}\ea$\\
\br
\end{tabular}
\end{indented}
\end{table}

\section{Embedding of (\ref{1.1}) with $b=2$ into a new $4$-component system admitting an $\mathfrak{sl}(3,\R)-$valued ZCR} \label{ZCR}

In this section we will consider the system (\ref{1.1}) with $b=2$, and rewrite it in the form

\bb\label{1.2*}
\begin{array}{l}
\left\{ \begin{array}{l}
D_{t}m_{1}+D_{x} \left( u_{1}\,u_2\,m_{1} \right)+\left(-u_1\,u_{2x}+2u_{2}u_{1x}\right)m_{1}=0,\\
\\
D_{t}m_{2}+D_{x} \left( u_{1}\,u_2\,m_{2} \right)+\left(-u_2\,u_{1x}+2u_{1}u_{2x}\right)m_{2}=0,\\
\end{array}\right.
\end{array}
\ee
where $u_1=u$, $u_2=v$ and $m_1=m$, $m_2=n$. 

In the paper \cite{geng} the authors already considered this system and they found that it admits a zero-curvature representation (ZCR) which depends on a parameter $\lambda$, $\lambda\in\mathbb{R}\setminus\{0\}$. Their ZCR is not $\mathfrak{sl}(3,\R)-$valued, nevertheless one can slightly modify their result and check that (\ref{1.2*}) also admits the $\mathfrak{sl}(3,\R)-$valued ZCR $ D_{t}X-D_{x}T+\left[X,T\right]=0$, with 

\[
X=\left(\begin{array}{ccc}
0 & \lambda m_{1} & 1\\
0 & 0 & \lambda m_{2}\\
1 & 0 & 0
\end{array}\right),
\]
and
\[
T=\left(\begin{array}{ccc}
\frac{1}{3\lambda^{2}}-u_{2} u_{1 x}\vspace{10pt}\quad & \frac{u_{1x}}{\lambda}-\lambda u_{1}u_{2} m_{1}\vspace{10pt}\quad & u_{1x}u_{2x}\vspace{10pt}\\
\frac{u_2}{\lambda}\vspace{10pt} & -\frac{2}{3\lambda^{2}}+u_{2} u_{1x}-u_{1}u_{2x}\vspace{10pt}\quad & -\lambda u_{1}u_{2} m_{2}+\frac{-u_{2x}}{\lambda}\vspace{10pt}\\
-u_{2}u_{1} & \frac{u_{1}}{\lambda} & \frac{1}{3\lambda^{2}}+u_{1}u_{2x}
\end{array}\right).
\]

Moreover, in the paper \cite{Li-Liu-Pop} has also been shown that (\ref{1.2*}) can be embedded in the following $4$-component Camassa-Holm
type hierarchy
\begin{equation} \label{li-Liu-Pop-sys}
\left\{ \begin{array}{l}
D_{t}m_{1}+D_{x}\left(\Gamma m_{1}\right)-\Gamma n_{2}+g_{1}g_{2}n_{2}+\left(f_{2}g_{2}+2f_{1}g_{1}\right)m_{1}=0,\vspace{5pt}\\
D_{t}m_{2}+D_{x}\left(\Gamma m_{2}\right)+\Gamma n_{1}-g_{1}g_{2}n_{1}-\left(f_{1}g_{1}+2f_{2}g_{2}\right)m_{2}=0,\vspace{5pt}\\
D_{t}n_{1}+D_{x}\left(\Gamma n_{1}\right)+\Gamma m_{2}-f_{1}f_{2}m_{2}-\left(f_{2}g_{2}+2f_{1}g_{1}\right)n_{1}=0,\vspace{5pt}\\
D_{t}n_{2}+D_{x}\left(\Gamma n_{2}\right)-\Gamma m_{1}+f_{1}f_{2}m_{1}+\left(f_{1}g_{1}+2f_{2}g_{2}\right)n_{2}=0,\vspace{5pt}
\end{array}\right.\label{eq:LLP_sys}
\end{equation}
where $m_{i}=u_{i}-u_{ixx}$, $n_{i}=v_{i}-v_{ixx}$, $i=1,2$, $f_{1}=u_{2}-v_{1x}$, $f_{2}=u_{1}+v_{2x}$, $g_{1}=v_{2}+u_{1x}$, $g_{2}=v_{1}-u_{2x}$ and $\Gamma$ is an arbitrary differentiable function of $u_{i},v_{i}$
and their partial derivatives with respect to $x$.  
As shown in \cite{Li-Liu-Pop}, system (\ref{eq:LLP_sys}) generalises several well known Camassa-Holm type equations and admits a ZCR too.

Like for \cite{geng}, also the ZCR originally considered in \cite{Li-Liu-Pop} for the system (\ref{eq:LLP_sys}) is not $\mathfrak{sl}(3,\mathbb{R})$-valued. However, one can check that an  $\mathfrak{sl}(3,\R)-$valued ZCR for (\ref{eq:LLP_sys}) is provided by 
\[
X=\left(\begin{array}{ccc}
0 & \lambda m_{1} & 1\\
\lambda n_{1} & 0 & \lambda m_{2}\\
1 & \lambda n_{2} & 0
\end{array}\right),
\]

\vspace{10pt}

\[
T=\left(\begin{array}{ccc}
\frac{1}{3\lambda^{2}}-f_{1}g_{1}\vspace{10pt}\quad & \frac{g_{1}}{\lambda}-\lambda\Gamma m_{1}\vspace{10pt}\quad & -g_{1}g_{2}\vspace{10pt}\\
-\lambda\Gamma n_{1}+\frac{f_{1}}{\lambda}\vspace{10pt} & -\frac{2}{3\lambda^{2}}+f_{1}g_{1}+f_{2}g_{2}\vspace{10pt}\quad & -\lambda\Gamma m_{2}+\frac{g_{2}}{\lambda}\vspace{10pt}\\
-f_{1}f_{2} & -\lambda\Gamma n_{2}+\frac{f_{2}}{\lambda} & \frac{1}{3\lambda^{2}}-f_{2}g_{2}
\end{array}\right),
\]
where $\lambda\in\mathbb{R}\setminus\{0\}$. 

The following result, which follows by direct computations, provides a generalisation of (\ref{eq:LLP_sys}) and hence a further generalisation of (\ref{1.2*}).

\begin{theorem}\label{teo3}
The four-component system
\begin{equation}\label{eq_gen}
\left\{ \begin{array}{l}
D_{t}m_{1}+D_{x}\left(\Gamma m_{1}\right)-\Gamma n_{2}+c_{1}\left(g_{1}g_{2}n_{2}+f_{2}g_{2}m_{1}+2f_{1}g_{1}m_{1}\right)\\
\\ \qquad-3c_{2}m_{1}-c_{3}n_{2}=0,\\
\\
D_{t}m_{2}+D_{x}\left(\Gamma m_{2}\right)+\Gamma n_{1}+c_{1}\left(-g_{1}g_{2}n_{1}-f_{1}g_{1}m_{2}-2f_{2}g_{2}m_{2}\right)\\
\\ \qquad+3c_{2}m_{2}+c_{3}n_{1}=0,\\
\\
D_{t}n_{1}+D_{x}\left(\Gamma n_{1}\right)+\Gamma m_{2}+c_{1}\left(-f_{1}f_{2}m_{2}-f_{2}g_{2}n_{1}-2f_{1}g_{1}n_{1}\right)\\
\\ \qquad+3c_{2}n_{1}+c_{3}m_{2}=0,\\
\\
D_{t}n_{2}+D_{x}\left(\Gamma n_{2}\right)-\Gamma m_{1}+c_{1}\left(f_{1}f_{2}m_{1}+f_{1}g_{1}n_{2}+2f_{2}g_{2}n_{2}\right)\\
\\ \qquad-3c_{2}n_{2}-c_{3}m_{1}=0,
\end{array}\right.\label{eq:gener_4_comp}
\end{equation}
where $m_{i}$, $n_{i}$, $f_{i}$, $g_{i}$ are given by
$$
\begin{array}{l}
m_{i}=u_{i}-u_{ixx},\qquad n_{i}=v_{i}-v_{ixx},\quad i=1,2,\vspace{5pt}\\
\\
f_{1}=u_{2}-v_{1x},\quad f_{2}=u_{1}+v_{2x},\quad g_{1}=v_{2}+u_{1x},\quad g_{2}=v_{1}-u_{2x},
\end{array}
$$
$c_{1},c_{2},c_{3}\in\mathbb{R}$ and $\Gamma$ is an arbitrary differentiable
function of $u_{i},v_{i}$ and their derivatives with respect to $x$,
admits the zero-curvature representation $D_{t}X-D_{x}T+\left[X,T\right]=0$ defined, for any $\lambda\in\mathbb{R}\setminus\{0\}$, by 
$$
X=\left(\begin{array}{ccc}
0 & \lambda m_{1} & 1\\
\lambda n_{1} & 0 & \lambda m_{2}\\
1 & \lambda n_{2} & 0
\end{array}\right)
$$
and 
$$
T=\left(\begin{array}{ccc}
c_{1}\left(\frac{1}{3\lambda^{2}}-g_{1}f_{1}\right)+c_{2}\vspace{10pt}\quad & c_{1}\frac{g_{1}}{\lambda}-\lambda m_{1}\Gamma\vspace{10pt}\quad & -c_{1}g_{1}g_{2}+c_{3}\vspace{10pt}\\
-\lambda n_{1}\Gamma+c_{1}\frac{f_{1}}{\lambda}\vspace{10pt} & c_{1}\left(-\frac{2}{3\lambda^{2}}+g_{1}f_{1}+g_{2}f_{2}\right)-2c_{2}\vspace{10pt}\quad & -\lambda m_{2}\Gamma+c_{1}\frac{g_{2}}{\lambda}\vspace{10pt}\\
c_{3}-c_{1}f_{1}f_{2} & -\lambda n_{2}\Gamma+c_{1}\frac{f_{2}}{\lambda} & c_{1}\left(\frac{1}{3\lambda^{2}}-g_{2}f_{2}\right)+c_{2}
\end{array}\right).
$$
\end{theorem}

\vspace{1cm}
System (\ref{li-Liu-Pop-sys}) is a particular instance of (\ref{eq_gen}), and in general they are not contact equivalent. In the particular case of (\ref{1.2*}) this fact readily follows from the following

\begin{theorem}
By choosing $v_{1}=v_{2}=0$ and $\Gamma=u_{1}u_{2}$, system $(\ref{eq_gen})$ reduces to the system
\bb\label{1.3}
\begin{array}{l}
\left\{ \begin{array}{l}
D_{t}m_{1}+D_{x} \left( u_{1}\,u_2\,m_{1} \right)+c_{1}\left(-u_1\,u_{2x}+2u_{2}u_{1x}\right)m_{1}-3c_{2}m_{1}=0,\\
\\
D_{t}m_{2}+D_{x} \left( u_{1}\,u_2\,m_{2} \right)+c_{1}\left(-u_2\,u_{1x}+2u_{1}u_{2x}\right)m_{2}+3c_{2}m_{2}=0,\\
\end{array}\right.
\end{array}
\ee
which is not contact equivalent to $(\ref{1.2*})$. In particular, $(\ref{1.3})$ reduces to $(\ref{1.2*})$ when $c_{1}=1,\,c_{2}=0$.

\end{theorem}
{\bf Proof:} The structure of the Lie algebra of classical symmetries of (\ref{1.3}) depends on $c_1$ and $c_2$. Indeed, by a direct computation one gets that the dimension of this Lie algebra is $4$ when $c_1\neq 1$, and $6$ when $c_1=1$. 
In particular, when $c_1\neq 1$ the Lie algebra is described by the characteristics $S_i=(\phi_{i},\psi_{i})$, with

$$\phi_{1}=u_x,\qquad \phi_{2}=u_t, \qquad \phi_{3}=t\left( 3c_{2} u-u_{t}\right), \qquad \phi_{4}=u,
$$
and
$$\psi_{1}=v_x,\qquad \psi_{2}=v_t,\qquad \psi_{3}=\left( -3c_{2} t v-tv_{t}-v\right),\qquad \psi_{4}=-v.
$$
In this first case the only non trivial Jacobi bracket is 

$$
\begin{array}{l}
\{S_2,S_3\}=-3c_{2}S_4+S_2.
\end{array}
$$
On the contrary, when $c_1=1$ the Lie algebra is described by the previous characteristics $S_i=(\phi_{i},\psi_{i})$ for $i=1,2,3,4$, and by two further characteristics $S_5=(\phi_{5},\psi_{5})$ and $S_6=(\phi_{6},\psi_{6})$ given by  

$$\phi_{5}=\e^{-2x}\left(u+u_x\right),\qquad \phi_{6}=\e^{2x}\left(-u+u_x\right),
$$
and
$$\psi_{5}=\e^{-2x}\left(v+v_x\right),\qquad \psi_{6}=\e^{2x}\left(-v+v_x\right).
$$
In this second case the only non trivial Jacobi brackets are 
$$
\begin{array}{l}
\{S_2,S_3\}=-3c_{2}S_4+S_2,\qquad \{S_1,S_5\}=2S_5,\qquad \{S_1,S_6\}=-2S_6,\qquad \{S_5,S_6\}=-4S_1.
\end{array}
$$
Thus the result follows by the invariance of symmetry algebras under contact transformations. 
$\square$

\section{Multi-peakons}\label{peak}

From now on, we assume that $b$ is a positive integer, and make the \"ansatz that system (\ref{1.1}) admits a superposition of peakon solutions of the form
\begin{equation}\label{4.1}
u(x,t)=\sum_{i=1}^Np_{i}\rme^{-|x-q_i|},\quad\quad v(x,t)=\sum_{i=1}^MP_{i}\rme^{-|x-Q_i|},
\end{equation}
where $N$ and $M$ are arbitrary positive integer numbers and $p_i,\,P_i,\,q_i$ and $Q_i$ are $2(N+M)$ smooth functions of $t$. We omitted the explicit dependence on $t$ for sake of simplicity. We shall denote derivative with respect to $t$ as $p^\prime_i,\,P^\prime_i,\,q^\prime_i$ and $Q^\prime_i$.

In the distributional sense, see \cite{schwartz} for further details, we have the following results:
\begin{equation}
\begin{array}{l}\label{4.2}
\ds{u_x=-\sum_{i=1}^N\sign{(x-q_i)p_i}\rme^{-|x-q_i|},\,\,v_x=-\sum_{i=1}^M\sign{(x-Q_i)P_i}\rme^{-|x-Q_i|}},\\
\\
\ds{u_{xx}=u-2\sum_{i=1}^Np_i\delta(x-q_i),\quad v_{xx}=v-2\sum_{i=1}^MP_i\delta(x-Q_i)}.
\end{array}
\end{equation}
Thus, one can write the momenta and their derivatives as
\bb\label{4.3}
\begin{array}{l}
\ds{m=2\sum_{i=1}^Np_i\delta(x-q_i),\,\, n=2\sum_{i=1}^MP_i\delta(x-Q_i),\,\, m_x=2\sum_{i=1}^Np_i\delta'(x-q_i)},\\
\\
\ds{n_x=2\sum_{i=1}^MP_i\delta'(x-Q_i),\,\,m_t=2\sum_{i=1}^N\left[p_i'\delta(x-q_i)-p_iq_i'\delta'(x-q_i)\right]},\\
\\
\ds{n_t=2\sum_{i=1}^N\left[P_i'\delta(x-Q_i(t))-P_iQ_i'\delta'(x-Q_i)\right].}
\end{array}
\ee

Hence, by substituting (\ref{4.1}), (\ref{4.2}) and (\ref{4.3}) into (\ref{1.1}), integrating against all pair of test functions with compact support and making use of the regularisation $\sign{(0)}=0$, one gets that the functions $p_i,\,P_i,\,q_i$ and $Q_i$ evolve according to the system of ODEs 
\bb\label{4.4}
\ba{lcl}
\ds{q_k'}&=&\ds{\left(\sum\limits_{m=1}^Np_me^{-|q_k-q_m|}\right)^{b-1}}\sum\limits_{j=1}^MP_je^{-|q_k-Q_j|},\quad 1\leq k\leq N,\\
\\

\ds{Q_\sigma'}&=&\ds{\left(\sum\limits_{m=1}^MP_m\rme^{-|Q_\sigma-Q_m|}\right)^{b-1}}\sum\limits_{j=1}^Np_j\rme^{-|Q_\sigma-q_j|},\quad 1\leq \sigma\leq M,
\\\\
\ds{p_k'}&=&\ds{p_k\left[(b+1)\sum\limits_{j=1}^N\sign{(q_k-q_j)p_j\rme^{-|q_k-q_j|}}\left(\sum\limits_{m=1}^MP_m\rme^{-|q_k-Q_m|}\right)^{b-1}\right.}\\
\\
&&\ds{\left.-(b-1)\sum\limits_{j=1}^N\sum\limits_{l=1}^M\sign{(q_k-q_j)}p_jP_l\rme^{-|q_k-q_j|-|q_k-Q_l|}\left(\sum\limits_{m=1}^Np_m\rme^{-|q_k-q_m|}\right)^{b-2}\right.}\\
\\
&&\ds{-\left.\sum\limits_{l=1}^M\sign{(q_k-Q_l)P_l\,\rme^{-|q_k-Q_l|}}\left(\sum\limits_{m=1}^Np_m\rme^{-|q_k-q_m|}\right)^{b-1}
\right]},\quad 1\leq k\leq N,\\
\\
\ds{P_\sigma'}&=&\ds{P_\sigma\left[(b+1)\sum\limits_{j=1}^M\sign{(Q_\sigma-Q_j)P_j\rme^{-|Q_\sigma-q_j|}}\left(\sum\limits_{m=1}^Np_m\rme^{-|Q_\sigma-q_m|}\right)^{b-1}\right.}\\
\\
&&\ds{\left.-(b-1)\sum\limits_{j=1}^M\sum\limits_{l=1}^N\sign{(Q_\sigma-Q_j)}P_jp_l\,\rme^{-|Q_\sigma-Q_j|-|Q_\sigma-q_l|}\left(\sum\limits_{m=1}^MP_m\rme^{-|Q_\sigma-Q_m|}\right)^{b-2}\right.}\\
\\
&&\ds{-\left.\sum\limits_{l=1}^N\sign{(Q_\sigma-q_l)p_l\,\rme^{-|Q_\sigma-q_l|}}\left(\sum\limits_{m=1}^MP_m\,\rme^{-|Q_\sigma-Q_m|}\right)^{b-1}
\right]}, \quad 1\leq \sigma\leq M.
\ea
\ee

Obtaining a solution of the system (\ref{4.4}) is in general a difficult task, however in the particular case when $N=M$, $q_j=Q_j$ and $p_j=P_j$, system (\ref{4.4}) reduces to
\bb\label{4.6}
\ba{lcl}
\ds{q'_k}&=&\ds{\left(\sum\limits_{j=1}^Np_j\rme^{-|q_k-q_j|}\right)^{b}},\\\
\\
\ds{p'_k}&=&\ds{p_k\sum\limits_{j=1}^N\sign{(q_k-q_j)p_j\rme^{-|q_k-q_j|}}\left(\sum\limits_{j=1}^Np_j\rme^{-|q_k-q_j|}\right)^{b-1}},
\ea
\ee
which is a result analogous to that obtained in \cite{anco}.

Although system (\ref{4.6}) shows the consistency of our results with those previously known, it also reflects a noteworthy difference between the scalar case considered in  \cite{anco} and the two component case of (\ref{1.1}). A particularly interesting manifestation of such differences is provided by the properties of $1$-peakon solutions of system (\ref{1.1}) which are discussed below. 

When $N=M=1$, after rearranging notations, system (\ref{4.4}) reads 
\bb\label{4.7}
\ba{l}
q'=p^{b-1}P\,\rme^{-|q-Q|},\quad Q'=P^{b-1}p\,\rme^{-|Q-q|}, \\
\\
 p'=-\sign{(q-Q)}\,\rme^{-|q-Q|}\,p^bP,\quad  P'=\sign{(q-Q)}\,\rme^{-|q-Q|}\,P^bp.
 \ea
\ee
Thus, in view of (\ref{4.7}), one has the following results.
\begin{theorem}\label{teo4}
Assume that $p,\,P,\,q$ and $Q$ are smooth functions satisfying $(\ref{4.7})$, then
\bb\label{4.8}
(q-Q)'=pP(p^{b-2}-P^{b-2})\,\rme^{-|q-Q|}
\ee
and
\bb\label{4.9}
(p\pm P)'=-\sign{(q-Q)}\,\rme^{-|q-Q|}\,pP(p^{b-1}\mp P^{b-1}).
\ee
\end{theorem}
{\bf Proof:} It follows from (\ref{4.7}) by a direct computation.$\square$

\begin{theorem}\label{teo5}
Assume that $p,\,P,\,q$ and $Q$ are smooth functions satisfying $(\ref{4.7})$, and assume that $\sign{(q-Q)}\neq0$,  then
\bb\label{4.10}
\left\{
\ba{lcl}
pP=\kappa,&\quad if\quad&b=2,\\
\\
\ds{\f{1}{p^{b-2}}+\f{1}{P^{b-2}}=\kappa,}&\quad if\quad&b\neq2,
\ea
\right.
\ee
where $\kappa$ is a constant.
\end{theorem}
{\bf Proof:}
It is enough to observe that $p'P^{b-1}=-p^{b-1}P'$.
$\square$

\begin{theorem}\label{teo6}
Assume that $p,\,P,\,q$ and $Q$ are smooth functions satisfying $(\ref{4.7})$, with $b\in \mathbb{N}$,
$b\neq2$ and $q=Q$,  then $q=\kappa^bt+x_0$, where $\kappa$ and $x_0$ are arbitrary constants. Moreover for any odd $b$ one has $p=P=\kappa$, whereas for any even $b$ one has $p=\pm\kappa$ and $P=\kappa$.
\end{theorem}

{\bf Proof:}
It follows from (\ref{4.8}) and the first equation of (\ref{4.7}).
$\square$

Theorem \ref{teo6} gives us a complete characterisation of peakons for $b\in \mathbb{N}$, $b\neq2$ and $q=Q$. Indeed, by rearranging notations, if $b$ is odd one has the solutions
\bb\label{4.11}
u(x,t)=c^{1/b}\rme^{-|x-ct-x_0|},\,\, v(x,t)=c^{1/b}\rme^{-|x-ct-x_0|},\quad b=1,3,5,\cdots.
\ee
On the other hand, if $b\neq 2$ is even, one has the following solutions
\bb\label{4.11}
u(x,t)=\pm c^{1/b}\rme^{-|x-ct-x_0|},\,\, v(x,t)=\pm c^{1/b}\rme^{-|x-ct-x_0|},\quad b=4,6,8,\cdots.
\ee

Notice that in the limit $u=v$ the system inherits the same solutions of (\ref{1.2}), however since $u$ and $v$ in ($\ref{4.11}$) do not need to have the same signal one also has the solutions $(u,v)=\pm(c^{1/b}e^{-|x-ct-x_0|},-c^{1/b}e^{-|x-ct-x_0|}).$

A more interesting situation occurs when $b=2$. Indeed, by Theorem \ref{teo4} and \ref{teo5}, one gets  $pP=k$ and $q=Q+x_0$, with $x_0$ a constant of integration. Therefore a straightforward integration of ODEs (\ref{4.7}) leads to the following
\begin{theorem}\label{teo7}
When $b=2$ the Cauchy problem for the system $(\ref{4.7})$, with the initial data $p(0)=p_0,\,P(0)=P_0,\,q(0)=q_0$ and $Q(0)=Q_0$, has the unique solution
\bb\label{4.14}
\ba{l}
q(t)=p_0\,P_0 \rme^{-|x_0|}t+q_0,\quad Q(t)=p_0\,P_0 \rme^{-|x_0|}t+Q_0,\\
\\
 p(t)=p_0\rme^{-t\,\sign{(x_0)}p_0\,P_0 \rme^{-|x_0|}},\,\, \ds{P(t)=P_0\,\rme^{t\,\sign{(x_0)} p_0\,P_0\rme^{-|x_0|}},}
\ea
\ee
where $x_0:=q_0-Q_0$.
\end{theorem}

In view of Theorem \ref{teo7}, for each fixed $t_0$ system (\ref{1.1}), with $b=2$, admits solutions with shape $\propto e^{-|x|}$. Actually, the Cauchy problem

\bb\label{4.15}
\begin{array}{l}
\left\{ \begin{array}{l}
m_{t}+3\,u_{x}\,v\,m+u^{b-1}\,v\,m_{x}=0,\\
\\
n_{t}+3\,v_{x}\,u\,n+v^{b-1}\,u\,n_{x}=0,\\
\\
 m=u-u_{xx},\quad n=v-v_{xx},\\
\\
u(x,0)=p_0{\rm e}^{-|x-q_0|},\quad v(x,0)=P_0{\rm e}^{-|x-Q_0|}
\end{array}\right.
\end{array}
\ee
admits a pair of 1-peakon solutions (that is, $N=M=1$ in (\ref{4.1})) given by
\bb\label{4.15}
\ba{l}
\ds{u(x,t)=p_0{\rm e}^{-t\,\sign{(x_0)}p_0\,P_0 \rme^{-|x_0|}}\,{\rm e}^{-|x-p_0\,P_0 {\rm e}^{-|x_0|}t-q_0|}},\\
\\
\ds{v(x,t)=P_0{\rm e}^{t\,\sign{(x_0)}p_0\,P_0 \rme^{-|x_0|}}\,{\rm e}^{-|x-p_0\,P_0 {\rm e}^{-|x_0|}t-Q_0|}}.
\ea
\ee

Let
\bb\label{4.16}
\ba{l}
\ds{u_0(x,t):=p_0\,{\rm e}^{-|x-p_0\,P_0\, t|}},\\
\\
\ds{v_0(x,t):=P_0\,{\rm e}^{-|x-p_0\,P_0\, t|}}.
\ea
\ee

A very quick calculation yields 
\bb\label{4.17}
\ba{l}
\|u(\cdot,t_0)\|_{L^p(\R)}= {\rm e}^{-p\,t_0\,\sign{(x_0)}p_0\,P_0 \rme^{-|x_0|}}\|u_0\|_{L^p(\R)},\\
\\
 \|v(\cdot,t_0)\|_{L^p(\R)}= {\rm e}^{p\,t_0\,\sign{(x_0)}p_0\,P_0 \rme^{-|x_0|}}\|v_0\|_{L^p(\R)},
 \ea
\ee
for each $t_0$ and $1\leq p\leq \infty$.

Recalling that $x_0=q_0-Q_0$, if $(q_0,Q_0)\rightarrow(0,0)$, then $x_0\rightarrow0$ and (\ref{4.15}) is equivalent to (\ref{4.16}).

\begin{figure}[!h]
         \includegraphics[width=0.3\textwidth]{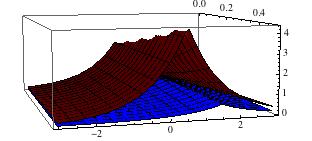}
          \includegraphics[width=0.3\textwidth]{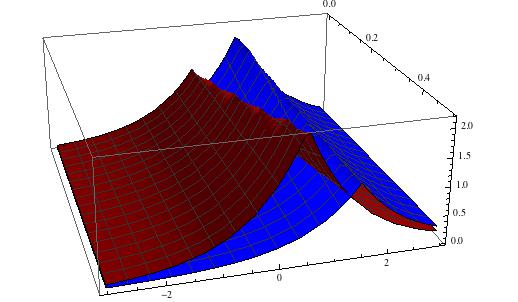}
           \includegraphics[width=0.3\textwidth]{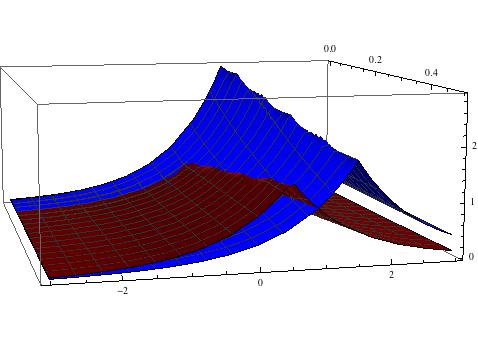}
        \caption{\small{The figures, from left to right, show the function (\ref{4.15}) with $\kappa=e$, $q_0=1$ and $p_0=1,\,2$, $3$, respectively. Function $u$ is represented in blue, whereas $v$ in red. The graphics were made with Mathematica by taking $(x,t)\in[0,1/2]\times[-3,3]$.}}
        \label{fig1}
\end{figure}

\begin{figure}[!h]
         \includegraphics[width=0.23\textwidth]{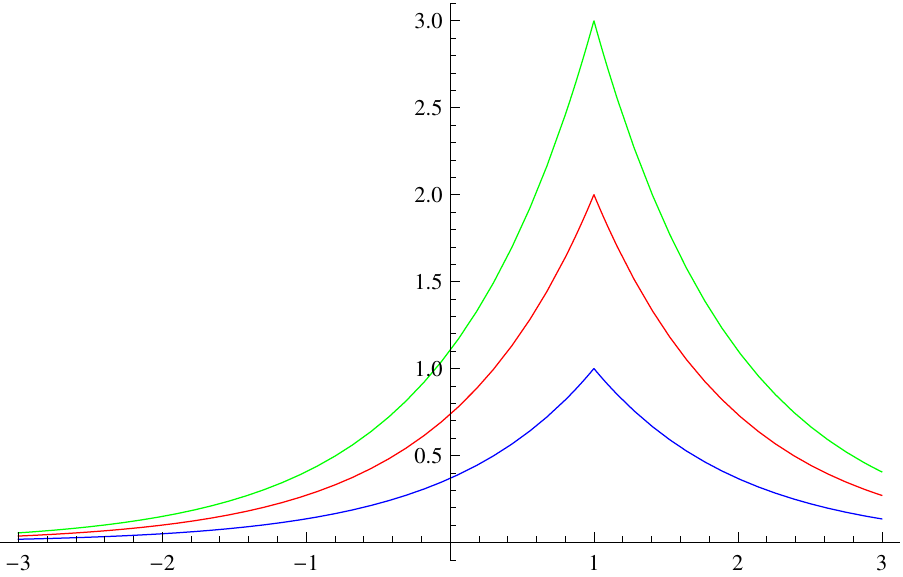}
          \includegraphics[width=0.23\textwidth]{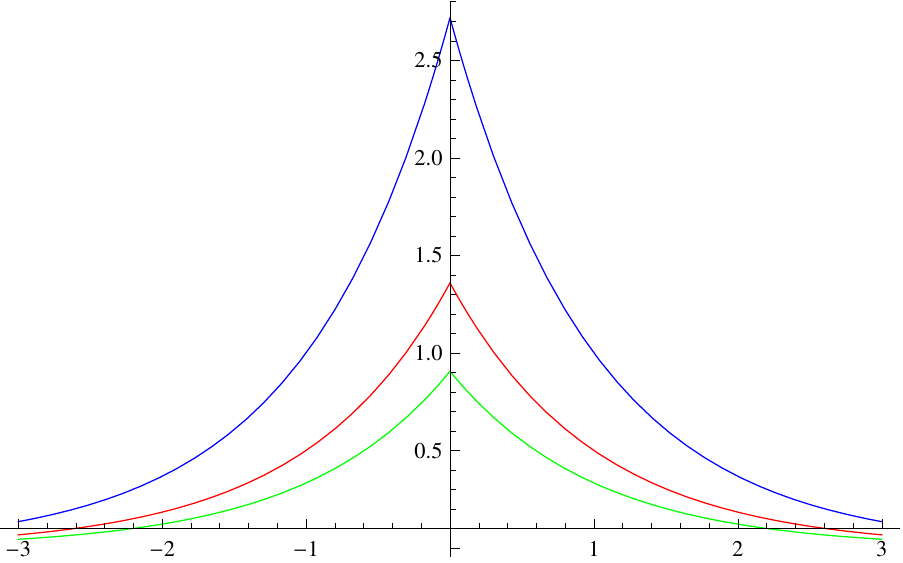}
          \includegraphics[width=0.23\textwidth]{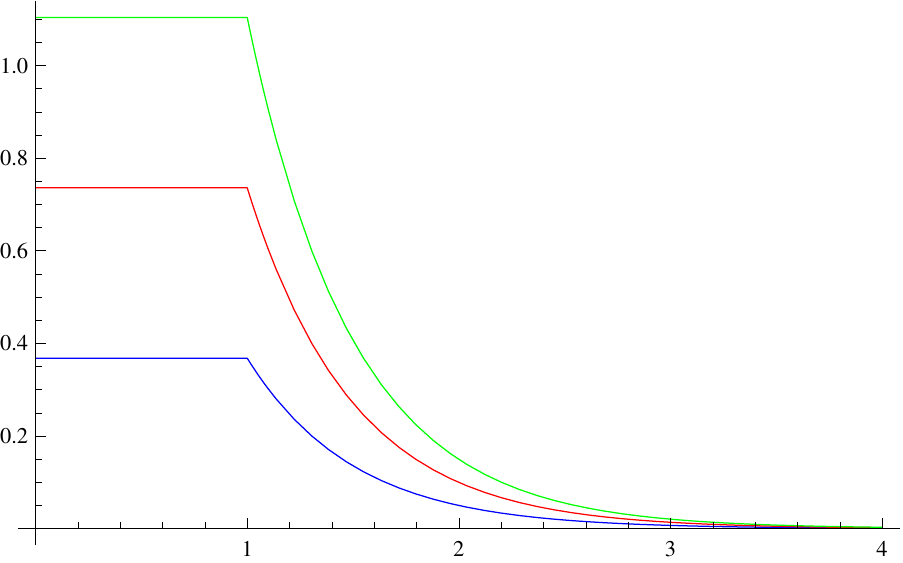}
          \includegraphics[width=0.23\textwidth]{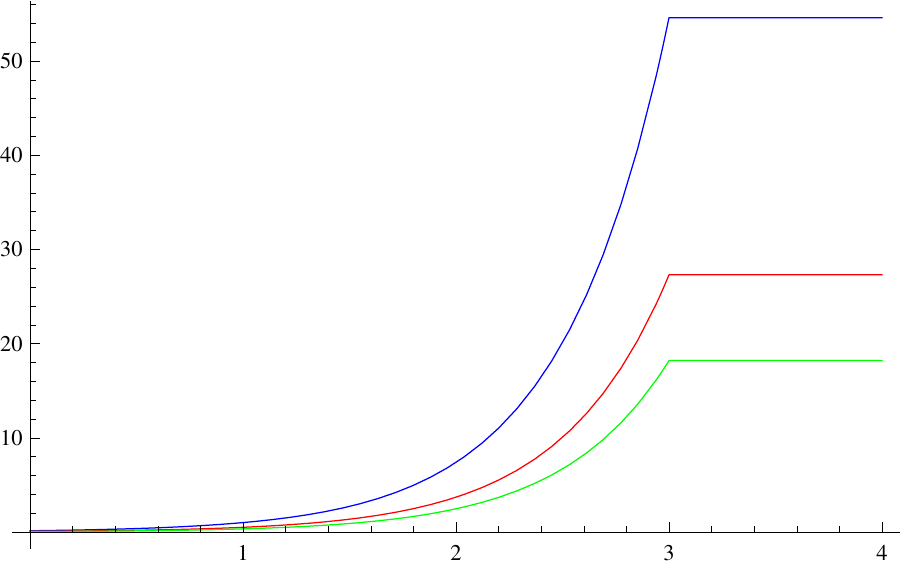}
          \caption{\small{The first and second figures, from left to right, describe $u(x,0)$ and $v(x,0)$ as provided by (\ref{4.15}) with $x\in[-3,3]$. The third and the fourth figures represent $u(2,t)$ and $v(2,t)$ for $t\in[0,4]$. In all cases $p_0\,P_0=\rme$ and $q_0=1$ and $Q_0=0$, whereas the values  $p_0=1,\,2$ and $3$ correspond to blue, red and green, respectively.}}
        \label{fig2}
\end{figure}

\section{Discussion}\label{dis}

Our results in this paper show that the multidimensional generalisation (\ref{1.1}) of the equation (\ref{1.2}) exhibits a behaviour slightly different from the scalar case when $b>2$, and very different for $b=1$ and $b=2$. Case $b=2$ is particularly interesting and richer.

From the point of view of Lie symmetries, when $b\neq2$ system (\ref{1.1}) has the same classical symmetry algebra of (\ref{1.2}), as follows by comparing with the results obtained in \cite{clark,anco}. On the contrary, when $b=2$, system (\ref{1.1}) admits a $6$-dimensional classical symmetry algebra which extends the $5$-dimensional symmetry algebra admitted by the Novikov equation \cite{bfi,pri3,anco}. For instance, when $b=2$ the system (\ref{1.2}) acquires the symmetry generator $X_4$ in (\ref{2.2}). Although at the level of point symmetries we have a minor change of behaviour of (\ref{1.1}) when compared with (\ref{1.2}), we begin to have better evidence of the differences when we look for higher order symmetries. In this direction, the only case we find such symmetries for system (\ref{1.1}) is just $b=2$, whereas equation (\ref{1.2}), being integrable for $b=1$ and $b=2$, has higher order symmetries for these cases, see \cite{camassa,how,nov}.

It has also been shown (see Theorem \ref{teo3}) that for  $b=2$ system (\ref{1.1}) can be embedded in a $4$-component system admitting an $\mathfrak{sl}(3,\R)-$valued zero-curvature representation which generalizes a $4$-component system found in \cite{Li-Liu-Pop}. Indeed, the $4$-component system described in our paper is not contact equivalent to that obtained in \cite{Li-Liu-Pop}.
 
With respect to conserved quantities, it is known that equation (\ref{1.2}) admits, for any positive integer value of $b$, the first integral (actually, a Hamiltonian for $b=1$ and $b=2$)
\bb\label{5.1}
{\cal H}_1[u]=\int_{\R}\left(u^2+u_{x}^2\right)dx,
\ee
see \cite{anco,camassa,bfi,pri1,pri2,how}. This integral corresponds to the Sobolev norm in $H^1(\R)$ of the solutions $u$ of (\ref{1.2}). Therefore, it is natural to expect that the bilinear form
\bb\label{5.2}
{\cal H}[u,v]=\int_{\R}\left(uv+u_{x}v_x\right)dx=\int_{\R}\left(vm+un\right)dx
\ee
would be a first integral of (\ref{1.1}). Notice that, for the scalar case, the integral (\ref{5.1}) can be derived by using the multiplier $u$ (in the sense of \cite{abeu1,abeu2,anco}) or the fact that (\ref{1.1}) is strictly self-adjoint (in the sense of \cite{ib2,ib6}). In the latter case, this first integral is derived from the scaling generator $X^b$ in (\ref{2.1}), see \cite{bfi,pri1,ib7}.

According to Table \ref{tabl1}, the first integral (\ref{5.2}) is derived as in the scalar case for $b=2$. However, (\ref{5.2}) is a first integral only for $b=2$. Indeed, by multiplying the first equation of system (\ref{1.1}) by $v$ and the second by $u$, simple manipulations yield the following relation:
\begin{equation}\label{5.3}
\ba{lcl}
D_t(vm+un)&=&-(b+1)[u_x\,v^b\,m+v_x\,u^b\,n]
\\
\\
&&-(u^{b-1}\,v^2\,m_x+v^{b-1}\,u^2\,n_x)+(v_t\,m+u_t\,n).
\ea
\end{equation}
Hence, (\ref{5.3}) provides a conservation law for (\ref{1.1}) only for  $b=2$, since just for this value of $b$  the right hand side of (\ref{5.3}) is a total derivative with respect to $x$. Indeed, by computing the variational derivative of the right hand side of (\ref{5.3}) one can see that this only happens for $b=2$. In this case (\ref{5.3}) can be rewritten as 

\begin{equation}\label{5.4}
\ba{l}
D_t(vm+un)+D_x[2u^2(2v^2-v_x^2-vv_{xx})\\
\\+u(4vu_xv_x-v_{tx}-2v^2u_{xx})+u_tv_x+u_xv_t-vu_{tx}-2v^2u_x^2]=0.
\ea
\end{equation}

Notice that in the degenerated case $u=v$ equation (\ref{5.3}) provides a conservation law for any value of $b$, since its right hand side is always a total derivative with respect to $x$.

Still about conserved quantities of (\ref{1.1}), in \cite{geng} it was shown that system (\ref{1.1}) with $b=2$ has a Hamiltonian. In \cite{lipla} and \cite{lijnmp} it was also found a second Hamiltonian and proved that (\ref{1.1}) with $b=2$ has a bi-Hamiltonian structure.

Finally, the most intriguing differences between the system (\ref{1.1}) and the scalar equation (\ref{1.2}) concern peakon solutions. Multi-peakon solutions of (\ref{1.1}) can be found by solving system (\ref{4.4}), which is in general a difficult task. However, 1-peakon solutions have been explicitly computed and show a noteworthy difference between 1-peakons of  (\ref{1.1}) and those of (\ref{1.2}). 

In the paper \cite{geng} the authors found the functions (\ref{4.16}) as solutions for system (\ref{1.1}) with $b=2$. These functions are $L^p(\R)-$integrables, for each $p$, and, in particular, are in $L^2(\R)$. However, we obtain a more general solution (\ref{4.15}) which admits (\ref{4.16}) as a particular case. These new solutions are $1$-peakons with non-constant amplitude and non-conservative norms in view of (\ref{4.17}), unless $x_0=0$, which implies that $q_0$ and $Q_0$ are just the same as well as $q(t)=Q(t)$, see Theorem \ref{teo7}.
 
 It is worth to notice that, if $x_0\neq0$ in (\ref{4.15}), then either one of the functions $u$ or $v$  blows up  when $t\rightarrow\infty$. This can be easily checked in the case $x_0=q_0>0,\,p_0\,P_0>0$ and $Q_0=0$, since for curve $t\mapsto x(t)=p_0P_0 e^{-|q_0|}t$ one has $|v(p_0P_0 e^{-|q_0|}t,t)|\rightarrow\infty$ when $t\rightarrow\infty$. This is another way to foresee that the norms of the solutions are not conserved. Particularly, they are not squared integrable solutions. However, in spite of this unboundedness, Theorem \ref{teo5} entails that the integral (\ref{5.2}) is bounded for these $1$-peakons since $uv,\,u_xv_x\thicksim {\rm e}^{-|x-p_0\,P_0\,t|}$.
  
\section{Acknowledgement}

The authors are grateful to P. L. da Silva for doing the figures of this paper and for useful discussions that motivated us to consider $N$ and $M$ multipeakons in (\ref{4.1}). The work of D. Catalano Ferraioli was partially supported by CNPq (grants no. $310577/2015-2$ and $422906/2016-6$). The work of I. L. Freire was partially supported by CNPq (grants no. $308516/2016-8$ and $404912/2016-8$). I. L. Freire would like to express his gratitude to the Departamento de Matemática -- UFBA, where this work begun, for the warm hospitality.


\section*{References} 
\begin{thebibliography}{99}

\bibitem{abeu1} S. Anco and G. Bluman, \newblock Direct construction method for conservation laws of partial differential equations. I. Examples of conservation law classifications,\newblock \emph{European J. Appl. Math.}, {\bf 13}, 545--566, (2002).

\bibitem{abeu2} S. Anco and G. Bluman, \newblock Direct construction method for conservation laws of partial differential equations. II. General treatment,\newblock \emph{European J. Appl. Math.}, {\bf 13}, 567--585, (2002).

\bibitem{anco} S. Anco, P. L. da Silva and I. L. Freire, \newblock A family of wave-breaking equations generalizing the Camassa-Holm and Novikov equations, \newblock \emph{J. Math. Phys.}, {\bf 56}, paper 091506, (2015).
 
\bibitem{bk} G. W. Bluman and S. Kumei,\newblock\emph{ Symmetries and Differential Equations}, Applied Mathematical Sciences 81, Springer, New York, (1989).

\bibitem{ba} G. W. Bluman and S. Anco, \newblock\emph{ Symmetry and Integration Methods for
Differential Equations}, Springer, New York, (2002).

\bibitem{2ndbook} G. Bluman, A. Cheviakov, S.C. Anco, \newblock Applications of Symmetry Methods to Partial Differential Equations, \newblock Springer Applied Mathematics Series 168, Springer, New York, (2010).
 
 \bibitem{bfi} Y. Bozhkov, I. L. Freire and N. H. Ibragimov,\newblock Group analysis of the Novikov equation, {\em Comp. Appl. Math.},  {\bf 33}, 193--202, (2014).
 
\bibitem{camassa} R. Camassa and D. D. Holm, \newblock An integrable shallow water equation with peaked solitons, \newblock \emph{Phys. Rev. Lett.}, {\bf 71}, 1661--1664, (1993).

\bibitem{clark} P. A. Clarkson, E. L. Mansfield and T. J. Priestley, \newblock Symmetries of a class of nonlinear third-order partial differential equations, {\em Math. Comput. Modelling.}, {\bf 25}, 195--212, (1997).

\bibitem{pri1}  P. L. da Silva and I. L. Freire, \newblock On certain shallow water models, scaling invariance and strict self-adjointness,  \newblock \emph{Proceeding Series of the Brazilian Society of Computational and Applied Mathematics}, (2015), DOI: /10.5540/03.2015.003.01.0022. See also, P. L. da Silva and I. L. Freire, {\it Strict self-adjointness and shallow water models}, e-print arXiv:1312.3992 (2013).  

\bibitem{pri2}  P. L. da Silva and I. L. Freire, \newblock An equation unifying both Camassa-Holm and Novikov equations,  \newblock \emph{Proceedings of the 10th AIMS International Conference}, (2015), DOI: 10.3934/proc.2015.0304.  

\bibitem{pri3} P. L. da Silva and I. L. Freire, \newblock On the group analysis of a modified Novikov equation, \newblock \emph{Interdisciplinary Topics in Applied Mathematics, Modeling and Computational Science}, {\bf117} \emph{Springer Proceedings in Mathematics and Statistics}, 161-166, (2015), DOI: 10.1007/978-3-319-12307-3$\_$23.



\bibitem{dehoho} A. Degasperis, D. D. Holm and A. N. W. Hone,\newblock A new integrable equation with peakon solutions, \newblock \emph{Theor. Math. Phys.}, {\bf133}, 1463--1474, (2002).


\bibitem{fokas1987} A. Fokas, Symmetries and integrability, Studies Appl. Math., {\bf 77}, 253--229, (1987).

\bibitem{geng} X. Geng and B. Xue, \newblock An extension of integrable peakon equations with cubic nonlinearity, \newblock \emph{Nonlinearity}, {\bf 22}, 1847--1856, (2009).

\bibitem{him1} A. A. Himonas and C. Holliman,  \newblock The Cauchy problem for the Novikov equation {\em Nonlinearity}, {\bf25}, 449--479, (2012).

\bibitem{him2} A. A. Himonas and J. Holmes,  \newblock Holder continuity of the solution map for the Novikov equation, {\em J. Math. Phys.}, {\bf54}, paper 061501, (2013).

\bibitem{him3} K. Grayshan and A. Himonas, \newblock Equations with peakon traveling wave solutions, {\em Adv. Dyn. Syst. Appl.}, {\bf 8}, 217--232, (2013).

\bibitem{him4} A. Himonas and C. Holliman, \newblock The Cauchy problem for a generalized Camassa-Holm equation, \newblock \emph{Adv. Differ. Equations}, {\bf 19}, 161--260, (2014).

\bibitem{him5} A. Himonas and D. Mantzavinos, \newblock An $ab$-family of equations with peakon traveling waves, {\em Proc. Amer. Math. Soc.}, {\bf 144}, 3797--3811, (2016). 

\bibitem{holm-staley} D. D. Holm and M. F. Staley, \newblock Wave structure and nonlinear balances in a family of evolutionary PDEs, \newblock \emph{Siam. J. Appl. Dyn. Sys.}, {\bf 2}, 323--380, (2003).

\bibitem{holm2010} D. D. Holm and R. I. Ivanov, \newblock Multi-component generalizations of the CH equations: geometrical aspects, peakons and numerical examples, \newblock \emph{J. Phys. A: Math. Theor.}, {\bf 43}, paper 492001, (2010)

\bibitem{how} A. N. W. Hone and J. P, Wang, \newblock Integrable peak on equations with cubic nonlinearities, \newblock \emph{J. Phys. A: Math. Theor.}, {\bf41}, 372002, 10 pp., (2008).


\bibitem{ib2} N. H. Ibragimov, \newblock A new conservation theorem, \newblock J. Math. Anal. Appl., {\bf 333}, 311--328, (2007).


\bibitem{ib6} N. H. Ibragimov,\newblock Nonlinear self-adjointness and conservation laws, {\em J. Phys. A: Math. Theor.}, {\bf44}, 432002, 8 pp., (2011).

\bibitem{ib7} N.H. Ibragimov, R.S. Khamitova, A. Valenti,\newblock Self-adjointness of a generalized Camassa-Holm equation, {\em Appl. Math. Comp.}, {\bf218}, 2579--2583, (2011).

\bibitem{i} N. H. Ibragimov, \newblock Transformation groups and Lie algebras, \newblock World Scientific, (2013).\newblock

\bibitem{kamp} P. H. van der Kamp and J. Sanders, Almost integrable evolution equations, Selecta Mathematica, {\bf 8}, 705--719, (2002).


\bibitem{lenjpa} J. Lenells, \newblock Conservation laws of the Camassa-Holm equation, {\em J. Phys. A: Math. Gen.}, {\bf 38}, 869--880, (2005).

\bibitem{lipla} N. Li and Q. P. Liu \newblock On bi-Hamiltonian structure of two-component Novikov equation, \newblock \emph{Phys. Lett. A}, {\bf 377}, 257--261, (2013).


\bibitem{lijnmp} H. Li, Y. Li and Y. Chen, \newblock Bi-hamiltonian structure of multi-component Novikov equation, \newblock \emph{J. Nonlin. Math. Phys.}, {\bf 21}, 509--520, (2014).


\bibitem{Li-Liu-Pop}N. Li, Q. P. Liu and Z. Popowicz, \newblock A four-component Camassa-Holm type hierarchy, {\em J. Geom. Phys.}, {\bf 85}, 29--39, (2014).

\bibitem{mik} A. V. Mikhailov, \newblock Introduction, Lect. Notes Phys., {\bf 767}, 1--18, (2009), DOI: 10.1007/978-3-540-88111-7$\_$0.

\bibitem{miknov} A. V. Mikhailov and V. S. Novikov, \newblock Perturbative symmetry approach, \newblock \emph{J. Phys. A: Math. Gen.}, {\bf 35}, 4775--4790, (2002).


\bibitem{nov} V. S. Novikov, \newblock Generalizations of the Camassa-Holm equation, \newblock \emph{J. Phys. A: Math. Theor.}, {\bf 42}, 342002, 14 pp., (2009).

\bibitem{ol}  P. J. Olver, \newblock Applications of Lie groups to differential equations, 2nd edition, Springer, New York, (1993).

\bibitem{qiao} Z. Qiao, \newblock A new integrable equation with cuspons and W/M-shape-peaks solitons, \newblock \emph{J. Math. Phys.}, {\bf 47}, paper 112701, (2006).

\bibitem{schwartz} L. Schwartz, \newblock Mathematics for the physical sciences, Dover, (2008) [English translation of L. Schwartz, \newblock M\'ethodes math\'ematiques pour les sciences physiques, (1966)].

 
\bibitem{vin} A. M. Vinogradov, Local symmetries and conservation laws, Acta Appl. Math., {\bf 2}, 21--78, (1984).




\end{thebibliography}
\end{document}